\documentclass[twocolumn]{aastex6}
\usepackage{amsmath}
\usepackage{amssymb}
\usepackage{graphicx}
\usepackage{subfigure}
\usepackage{natbib}
\usepackage{enumitem}


\newcommand{\msun}{\, \mbox{M}_{\odot}}
\newcommand{\muG}{\, \mu\mbox{G}}
\newcommand{\flash}{{\normalfont\scshape flash }}
\newcommand{\paramesh}{{\normalfont\scshape paramesh }}

\begin{document}
\title{The Evaporation and Survival of Cluster Galaxies' Coronae. Part I: The Effectiveness of Isotropic Thermal Conduction Including Saturation}
\author{Rukmani Vijayaraghavan\altaffilmark{1,2}}
\author{Craig Sarazin\altaffilmark{1}}
\email{rukmani@virginia.edu}
\altaffiltext{1}{Department of Astronomy, University of Virginia, 530 McCormick Rd., Charlottesville, VA 22904, USA}
\altaffiltext{2}{NSF Astronomy \& Astrophysics Postdoctoral Fellow}        

\begin{abstract}
We simulate the evolution of cluster galaxies' hot interstellar medium (ISM) gas due to ram pressure and thermal conduction in the intracluster medium (ICM). At the density and temperature of the ICM, the mean free paths of ICM electrons are comparable to the sizes of galaxies, therefore electrons can efficiently transport heat due to thermal conduction from the hot ICM to the cooler ISM. Galaxies consisting of dark matter halos and hot gas coronae are embedded in an ICM-like `wind tunnel' in our simulations. In this paper, we assume that thermal conduction is isotropic and include the effects of saturation.  We find that as heat is transferred from the ICM to the ISM, the cooler denser ISM expands and evaporates. This process is significantly faster than gas loss due to ram pressure stripping; for our standard model galaxy the evaporation time is $160$ Myr while the ram pressure stripping timescale is $2.5$ Gyr. Thermal conduction also suppresses the formation of shear instabilities, and there are no stripped ISM tails since the ISM evaporates before tails can form. Observations of long-lived X-ray emitting coronae and ram pressure stripped X-ray tails in galaxies in group and cluster environments therefore require that thermal conduction is suppressed or offset by some additional physical process. The most likely process is anisotropic thermal conduction due to magnetic fields in the ISM and ICM, which we simulate and study in the next paper in this series.
\end{abstract}

\maketitle

\section{Introduction}
\label{sec:intro}

Galaxies in cluster environments are gas poor \citep[e.g.,][]{Davies73,Haynes84,Solanes01,Boselli06,Cortese11,Jaffe16} and consequently, on average, have significantly lower star formation rates and higher early type fractions than galaxies in the field \citep{Dressler80,Kennicutt83,Postman84,Zabludoff98,Lewis02,Gomez03,Weinmann06,Rasmussen12, Haines13}. In these dense environments with deep potential wells, galaxies are stripped of dark matter, stars, and gas through tidal forces and harassment \citep{Moore96,Moore98,Gnedin03a,Gnedin03b,Mastropietro05,Villalobos12,Villalobos14}. The hot X-ray emitting intracluster medium (ICM), with temperatures $T_{\rm ICM} \sim 10^7 - 10^8$ K and densities $n_e \sim 10^{-3}$ cm$^{-3}$, exerts ram pressure on the interstellar medium (ISM) gas in galaxies \citep{Gunn72,Quilis00}. Ram pressure causes a drag force that may remove the hot coronal / halo component of the ISM \citep{Larson80,Kawata08,McCarthy08,Shin14,Roediger15a,Roediger15b}, as well as the cold molecular and atomic ISM \citep{Schulz01,Vollmer01,Roediger06,Kapferer09,Tonnesen09,Ruszkowski14,Tonnesen14}. In addition, thermal conduction can transfer heat from the ICM to the ISM, resulting in the evaporation of the hot ISM \citep{Cowie77b,Sarazin86}. 

On the other hand, X-ray observations of cluster galaxies reveal that a significant  fraction of cluster galaxies have compact X-ray emitting coronae. \citet{Vikhlinin01} observed the two central galaxies in the Coma cluster with \textit{Chandra}, and detected compact ($\sim 3$ kpc) X-ray coronae of 1-2 keV centered on these galaxies.
These coronae were studied in more detail by \citet{SFS+14}.
Later studies extended these observations to other central \citep{Yamasaki02} and  satellite galaxies in clusters \citep[e.g.,][]{Sun05,Sun05c}. Systematic studies by \citet{Sun07} show that  $\sim 60\%$ of $L_K > 2 L_*$ galaxies in rich clusters have compact X-ray emitting coronae. \citet{Jeltema08} show that an even larger fraction of group galaxies have X-ray coronae, up to $80\%$ of $L_K > L_*$ galaxies. The frequency of these detections in group and cluster galaxies imply that galactic coronae survive for timescales on the order of many orbital times in clusters. These observations contradict theoretical expectations for gas retention timescales in cluster galaxies subject to ram pressure stripping and evaporation due to thermal conduction.

The rate at which a galaxy is stripped of its gas by ram pressure ($P_{\rm ram} = \rho_{\rm ICM} v_{\rm galaxy}^2$) is determined by the galaxy's orbit within the cluster, including the orbital ellipticity, the pericentric distance to the cluster center which affects both the local $\rho_{\rm ICM}$ encountered by the galaxy and the maximum orbital velocity, and the average orbital velocity of the galaxy. A galaxy's' resistance to ram pressure is driven by its internal thermal pressure, which, assuming hydrostatic equilibrium, is determined by its density profile and gravitational potential; one expects that more massive galaxies are more resistant to ram pressure. Using simulations of a cluster with a realistic population of initially hot ISM gas-rich galaxies,\citet{Vijayaraghavan15b} showed  that even relatively massive galaxies ($M > 10^{11} \msun$) lose $90\%$ of their gas mass within 2.5 Gyr over a range of galaxy orbits and velocities; lower mass galaxies lose all their gas within this period. Massive galaxies lose all of their gas within $4 - 5$ Gyr. These timescales are an upper limit for the removal of the hot coronal ISM, since in these simulations, the coronae / hot halos in galaxies extend out to galaxies' virial radius and comprise $10 \%$ of the galaxies' mass. Ram pressure alone can therefore efficiently remove the hot coronal gas within one dynamical time ($t_{\rm dynamical} \sim 2-3$ Gyr for rich clusters). 

Thermal conduction, the transfer of heat via electrons and ions, affects the evolution of the ICM in various ways. The transfer of heat from the hotter outer ICM to the cooler core,  suppressing cooling and star formation, has been proposed as a partial solution to the cooling flow problem in clusters \citep[e.g.,][]{Bertschinger86,Narayan01,Zakamska03,Voigt04,Jubelgas04,Ghizzardi04,Wagh14}. A caveat with this process is that thermal conduction can be suppressed by ICM magnetic fields to values significantly below the classical Spitzer value  \citep{Spitzer62}. We explore the particular effects of anisotropic thermal conduction due to magnetic fields
on galactic coronae in the next paper in this series (Paper II).  Thermal conduction has also been proposed as a potential mechanism to transport heat generated by central AGN to the outer regions of clusters (reviewed in  \citealt{McNamara07}). Using numerical simulations, \citet{Dolag04} showed  that thermal conduction can result in an isothermal ICM core, but is less effective in cool core clusters. \citet{Smith13} also studied the effect of isotropic thermal conduction on ICM temperature profiles, finding that including conduction resulted in more isothermal cores and a thermally uniform ICM. \citet{Voit11} suggest that radiatively cooled gas in cluster cores can trigger AGN feedback when thermal conduction cannot balance cooling. 

At typical temperatures of the ICM and galactic coronae, thermal conduction can be particularly effective in transporting heat from the ICM to ISM. The mean free path of ICM electrons is $\sim 10$ kpc, comparable to the sizes of galactic coronae. Assuming isotropic thermal conduction, ICM electrons can  transport heat between the ISM and ICM \citep{Spitzer53,Sarazin86}. The expected timescale for this process assuming steady state evaporation and mass loss  for a uniform constant density ($\bar{n}$) sphere of radius $R$ (in kpc), is $t_{\rm evap} = 3.3 \times 10^{20} \bar{n} R^2 T^{-5/2} (\ln \Lambda / 30) \,  \rm{Myr}$ \citep{Cowie77}. $t_{\rm evap} \sim 10^7$ years for a typical galaxies' parameters, significantly shorter than the ram pressure stripping timescale of $\sim 3 \times 10^9$ years, although this assumption does not include the effects of a gravitational potential and a density and temperature gradient. \citet{Cowie77b} also calculated the evaporation timescales of galaxies embedded in $10^8$ K gas;  their calculated timescale is proportional to the size of coronae, but is significantly shorter than ram pressure stripping times and mass loss replenishment from stars. A combination of thermal conduction, which evaporates the ISM and results in a more uniform temperature distribution, ram pressure stripping that removes any remnant ISM gas, and tidal stripping that decreases the depth of the galaxies' potential well should therefore efficiently remove hot galactic coronae.

Gas removal from cluster galaxies must therefore be offset by physical mechanisms that shield against stripping and conduction, or replenish gas loss from galaxies. The ICM is threaded by $\muG$ magnetic fields (e.g., as reviewed in \citealt{Carilli02}; \citealt{Govoni04}; \citealt{Kronberg05}; \citealt{Ryu12}), which can suppress shear instabilities that are precursors to the dissipation of the ISM at the the ISM-ICM interface. However, the magnetic pressure in the ICM is $\sim 10^2$ times weaker than the thermal and ram pressure and therefore does not quantitatively affect the hydrodynamics of the stripping processes \citep{Shin14,Tonnesen14,Ruszkowski14,Vijayaraghavan16}. On the other hand, magnetic fields will affect the rate of ISM evaporation due to thermal conduction. In the presence of a local magnetic field, electrons will gyrate around the magnetic field \citep{Spitzer62}. The gyroradius of a typical electron ($r_g = m v_{\perp} c / q B$) is $\sim 2 \times 10^{8}$ cm, significantly smaller than electron mean free paths. Electrons therefore cannot cross magnetic field lines, and are constrained to only transport heat along the direction of the magnetic field. This anisotropic thermal conduction can drastically reduce the effectiveness of evaporation due to thermal conduction. A viscous ICM will also suppress shear instabilities \citep{Roediger13,Roediger15b}. Magnetic fields and viscosity therefore qualitatively affect the stripping of galaxies, resulting in more coherent, long-lived stripped tails, but do not dramatically affect the gas loss rate of galactic coronae due to stripping. 

In this paper (Part I) and the following paper on anisotropic thermal conduction in cluster galaxies, we simulate galaxies embedded in wind tunnel simulation boxes with ICM-like physical properties.
In Part I, we describe the process of gas loss due to isotropic thermal conduction including saturation, in the absence of magnetic fields. We quantify the gas loss rate of galaxies due to thermal conduction and ram pressure stripping, the evolution the effective heat flux as the galaxy evaporates, and the effectiveness of thermal conduction on varying galaxy mass and ICM densities. In Part II, we describe the effect of magnetic fields and subsequently anisotropic thermal conduction on stripping and evaporation over a range of magnetic field configurations that either shield the galaxy or connect the galaxy and the ICM, the longevity of stripped tails, the relationship between tail morphology and magnetic field configuration, and the timescale over which gas evaporates due to anisotropic thermal conduction. 

\section{Simulation Methods}
\label{sec:methods}

We use the \flash 4.3 code \citep{Fryxell00,Dubey08}, a parallel $N$-body plus Eulerian hydrodynamics code with adaptive mesh refinement (AMR), to perform the simulations in this paper. Our simulations consist of galaxies with dark matter and gas components in a wind tunnel of ICM gas. In our simulations, massive particles are mapped to the grid using cloud-in-cell (CIC) mapping. The potential on the mesh is calculated from the total dark matter plus gas density using the direct multigrid solver \citep{Ricker08}. AMR is implemented on the Eulerian  mesh in \flash using \paramesh \citep{Macneice00}. We use the directionally split piecewise-parabolic method (PPM) solver in \flash to evolve Euler's equations of hydrodynamics. Thermal conduction is explicitly implemented by calculating the heat flux on the boundaries of each cell, as described below, and the subsequent change in internal energy for a given cell. The computational timestep is diffusion limited to ensure stability.\footnote{$\Delta t = 0.5 \frac{\Delta x^2}{D}$, with $D = \kappa / \rho c_v$, where $\kappa$ is the conductivity, $\rho$ is the gas density, and $c_v$ is the specific heat capacity.}

\subsection{Theoretical Background: Evaporation of Coronae in a Hot ICM}
\label{sec:methods_evap}

Cool galactic coronae embedded in hot intracluster media form temperature gradients. In the absence of magnetic fields and any other heat sources, the temperature gradient will result in the flow of heat, primarily through electrons, from the hot ICM to the cooler galactic ISM (\citealt{Spitzer53}, \citealt{Spitzer62}). As a result, the galactic ISM gas will expand and evaporate into the ICM. Quantitatively, the flow of heat is expressed in the form of the heat flux, $\mathbf{Q}$, where
\begin{equation}
\mathbf{Q} = -\kappa \mathbf{\nabla} T_e \, ,
\end{equation}
with
\begin{equation}
\kappa = \epsilon \delta_{T} \kappa_{\rm{Sp}} \, ,
\end{equation} 
where $\mathbf{\nabla} T_e$ is the temperature gradient, $\kappa$ is the conduction coefficient, and $\epsilon = 0.419$ and $\delta_T = 0.225$ for a hydrogen ($Z = 1$) plasma. $\kappa_{\textrm{Sp}}$, the Spitzer thermal conductivity,  is:
\begin{equation}
\kappa_{\textrm{Sp}} = 20 \left(\frac{2}{\pi}\right)^{3/2} \frac{k_B^{7/2} T_e^{5/2}}{m_e^{1/2} q_e^{4} Z \ln \Lambda } \, .
\end{equation}
Here, $k_B$ is the Boltzmann constant, $T_e$ is the electron temperature, $m_e$ and $q_e$ are the mass and charge of an electron, and $\ln \Lambda$ is the Coulomb logarithm,
\begin{equation}
\ln \Lambda = 37.8 + \ln \left[\left(\frac{T_e}{10^8 \textrm{K}}\right) \left(\frac{n_e}{10^{-3} \textrm{cm}^{-3}}\right)\right] \, .
\end{equation}
Thermal conduction in these media is primarily carried out through electrons, since $\kappa_{\textrm{Sp}} \propto m_e^{-1/2}$. The electron mean free path is \citep{Spitzer62,Sarazin86}:
\begin{equation}\label{eqn:lambdamfp}
\lambda_e = \frac{3^{3/2} (k_B T_e)^2}{4 \pi^{1/2} n_e q_e^{4} \ln \Lambda} \simeq 22~\textrm{kpc} \left(\frac{T_e}{10^8 \textrm{K}}\right)^2 \left(\frac{n_e}{10^{-3} \textrm{cm}^{-3}} \right)^{-1} \, .
\end{equation}
The Spitzer thermal conductivity can therefore be written as:
\begin{equation}
\kappa_{\textrm{Sp}} = 13.8 ~ n_e \lambda_{e} \frac{k_B^{3/2} T_e^{1/2}}{m_e^{1/2}}
\, {\rm erg} \, {\rm cm}^{-1} \, {\rm s}^{-1} \, {\rm K}^{-1} \, .
\end{equation}

The heat flux, $\mathbf{Q}$, is inversely proportional to the length scale of the temperature gradient, $l_T$; the smaller the value of $l_T$, the higher the heat flux. However, \citet{Cowie77} show that since the rate at which heat is transported is limited by the average electron speed in a given medium, the effective heat flux  can saturate.
They find that the magnitude of the saturated heat flux when $\lambda_e \gg l_T$ is:
\begin{equation}
Q_{\textrm{sat}} = 0.4 \left(\frac{2 k_B T_e}{\pi m_e} \right)^{1/2} n_e k_B T_e.
\end{equation}
The effective heat flux is obtained by interpolating between $\mathbf{Q}$ and $\mathbf{Q_{\textrm{sat}}}$:
\begin{equation}
\frac{1}{\mathbf{Q_{\textrm{eff}}}} = \frac{1}{\mathbf{Q}}  + \frac{1}{\mathbf{Q_{\textrm{sat}}}} .
\end{equation}
In terms of the heat flux ratio, $\sigma = Q / Q_{\textrm{sat}}$, the effective heat flux is:
\begin{equation}
\mathbf{Q_{\textrm{eff}}} = \frac{- \kappa \mathbf{\nabla} T_e}{1 + \sigma},
\end{equation}
which alternatively reduces to 
\begin{equation}
\mathbf{Q_{\textrm{eff}}} = \frac{\kappa T_e}{l_T + 4.2\lambda_e} \frac{\mathbf{\nabla} T_e}{| \mathbf{\nabla} T_e|}.
\end{equation}
When $l_T \gg \lambda_e $, the effective heat flux obeys the classical form, while in the regime of $l_T \ll \lambda_e $, the effective heat flux is saturated. For typical galactic coronae in cluster environments, $l_T \simeq \lambda_e $ since the sizes of typical galactic coronae are comparable to the electron mean free paths in equation~\ref{eqn:lambdamfp}. Thermal conduction in these regions is therefore at least partially saturated.

\subsection{Initial Conditions}
\label{sec:methods_initial}

We perform a series of `wind-tunnel' simulations of a galaxy embedded in an ICM-like wind. These simulations, in the rest frame of the galaxy, consist of a galaxy in the center of a  cubic box. One side of the box is an inflow boundary, though which the ICM gas flows in at a constant velocity. The other five sides of the box are outflow boundaries. The density, pressure, and temperature of the ICM wind are initially constant throughout the box. The center of the galaxy is at the center of the box, to prevent the development of an artificial potential gradient within the box from an asymmetric mass distribution of the ICM wind with respect to the galaxy.

The galaxy consists of a spherically symmetric dark matter potential and hot ISM gas in hydrostatic equilibrium with the gravitational potential. The galaxy's collisionless dark matter component and gas are initialized using the procedure described in \citet{Vijayaraghavan15b}, which we summarize here. The galaxy's total density profile is: 
\begin{equation}
\rho_{\rm tot} = \rho_{\rm DM} + \rho_{\rm ISM} + \rho_{\rm ICM,background}.
\end{equation}     
This total density profile is described by a Navarro-Frenk-White (NFW, \citealt{Navarro97}) profile:
\begin{equation}
  \rho_{\rm tot}(r \leq R_{200}) = \frac{\rho_{\rm s}}{r/r_{\rm s} (1 + r/r_{\rm s})^2} 
\end{equation}
where $R_{200}$ is the galaxy's virial radius, and $r_s$ and $\rho_s$ are the scale radius and scale density parameters.
$\rho_{\rm ICM,background}$ is the gas density beyond the ISM and is initially constant, while $\rho_{\rm ISM} $ is assumed to be a singular isothermal sphere distribution with:
\begin{equation}
\rho_{\rm ISM} (r \leq R_{200}) = \frac{\rho_0 r_0^2 }{r^2}.
\end{equation}
We assume that the gas mass fraction of the ISM is 10\% of the total mass of the galaxy, not including the background ICM. At $r >  R_{200}$, the dark matter is assumed to have an exponential fall off in density and the density of gas is the constant ICM background density. Given the dark matter density profile, $\rho_{\rm DM} = \rho_{\rm tot} - \rho_{\rm ISM} - \rho_{\rm ICM,background}$, we initialize the positions and velocities of  massive dark matter particles to determine the galaxy's gravitational potential as described in \citet{Vijayaraghavan15b}, using the procedure outlined in \citet{Kazantzidis04}. 

The galaxy's pressure profile is initialized assuming hydrostatic equilibrium, where $\rho_{\rm gas} = \rho_{\rm ISM} + \rho_{\rm ICM,background}$ and $\Phi$ is the total gravitational potential:
\begin{equation}
\dfrac{dP}{dr} = -\rho_{\rm gas}\dfrac{d\Phi}{dr}.
\end{equation}  
The temperature $T$ is calculated from the usual ideal gas law:
\begin{equation}
P = \frac{k_B}{\mu m_p} \rho_{\rm gas} T,
\end{equation}
where $k_B$ is the Boltzmann constant, $\mu \simeq 0.59$ is the mean molecular weight for a fully ionized hydrogen--helium plasma, and $m_p$  is the proton mass. We assume a constant $\gamma = 5/3$ equation of state throughout. The density, temperature, and pressure of the galaxy are continuous with the background ICM at $r = R_{200}$. In our base series of simulations, we adopt $\rho_{\rm ICM} = 2 \times 10^{-27} $ g cm$^{-3}$, $P_{\rm ICM} = 2 \times 10^{-11}$ erg  cm$^{-3}$, and $T_{\rm ICM} = 7.14 \times 10^7$ K. The mass of the galaxy is $M_{200} = 2.8 \times 10^{11} \msun$ and its virial radius is $R_{200} = 92.16$ kpc and its scale radius is $R_{\rm s} = 18.43$ kpc. We also ran a simulation with a lower density ICM, with $\rho_{\rm ICM} = 2 \times 10^{-28} $ g cm$^{-3}$, $P_{\rm ICM} = 2 \times 10^{-12}$ erg  cm$^{-3}$, and $T_{\rm ICM} = 7.14 \times 10^7$ K, and of a higher mass galaxy with $M_{200} = 2.8 \times 10^{12} \msun$ with a virial radius of $R_{200} = 190$ kpc. The galaxy properties for both the standard galaxy and the higher mass case were chosen from the galaxy sample in \citet{Vijayaraghavan15b}, where they are discussed further. We note that the galaxy scale radius and gas density profile slope don't have a significant effect on ram pressure stripping \citep{Vijayaraghavan15b}.

\section{Results}
\label{sec:results}

\subsection{No Evaporation, with Ram Pressure}
\label{sec:results_nocond}

\begin{figure*}[!htbp]
  \begin{center}
{\includegraphics[width=\textwidth]{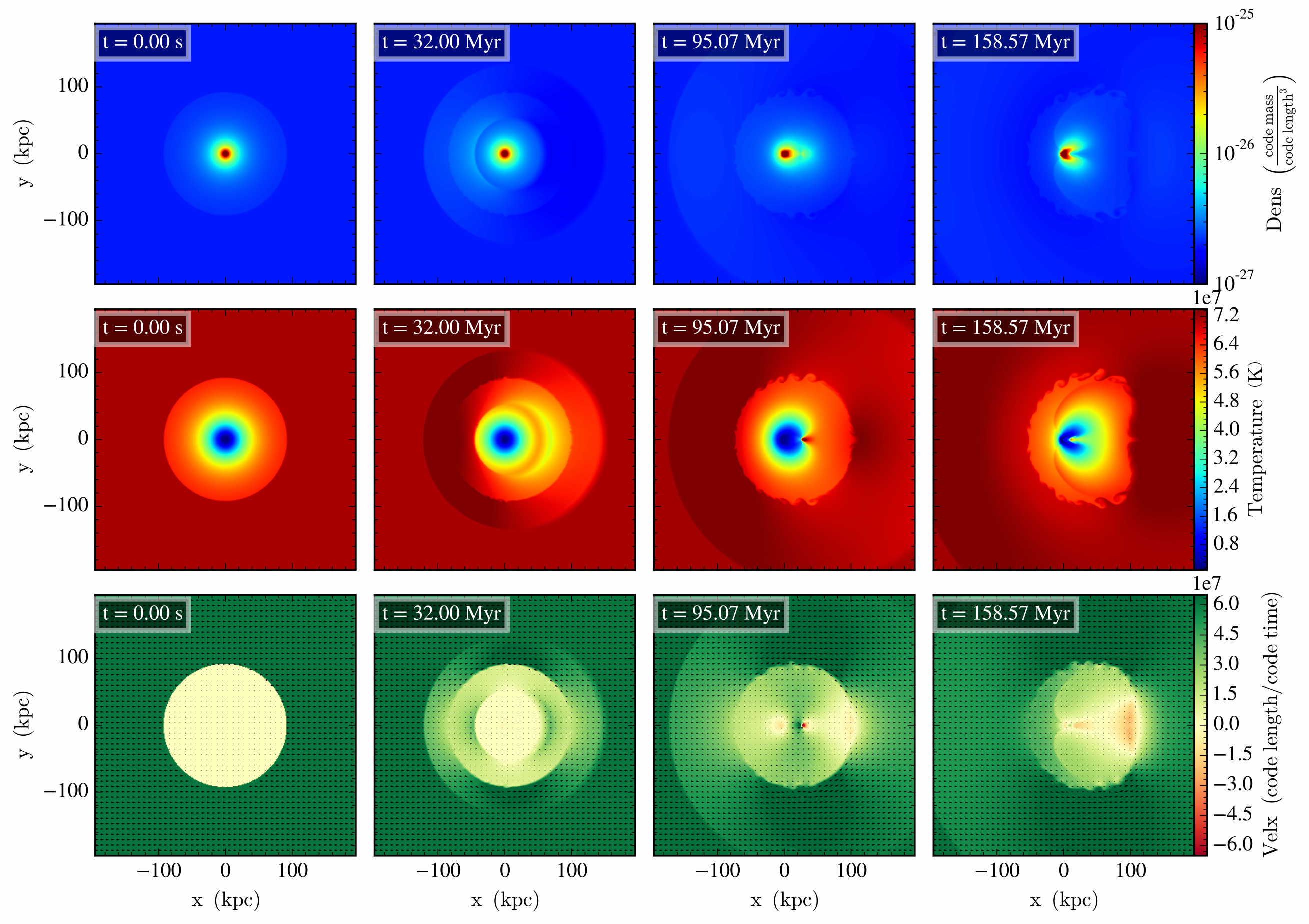}\label{fig:denstempvelxwind1}}
    \caption{Slices of density, temperature, and the $x$ component of the velocity in a simulation of gas loss due to ram pressure with no thermal conduction from  $t = 0 - 160$ Myr. The velocity slices are annotated with velocity vectors.
    \label{fig:slicenocond}}

\end{center}  
\end{figure*}

\begin{figure*}[!htbp]
  \begin{center}
    {\includegraphics[width=\textwidth]{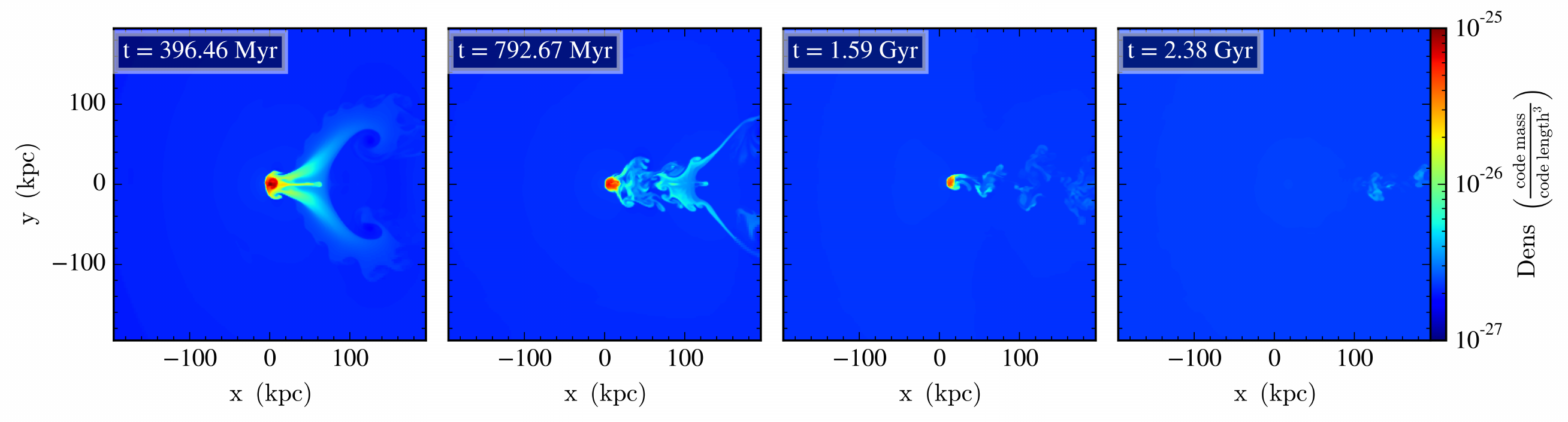}\label{fig:denswind2}}
    \caption{Slices of density in simulation of gas loss due to ram pressure with no thermal conduction from  $t = 400 - 2400$ Myr.
    \label{fig:slicenocondlate}}

\end{center}  
\end{figure*}

\begin{figure}[!htbp]         
    {\includegraphics[width=0.4\textwidth]{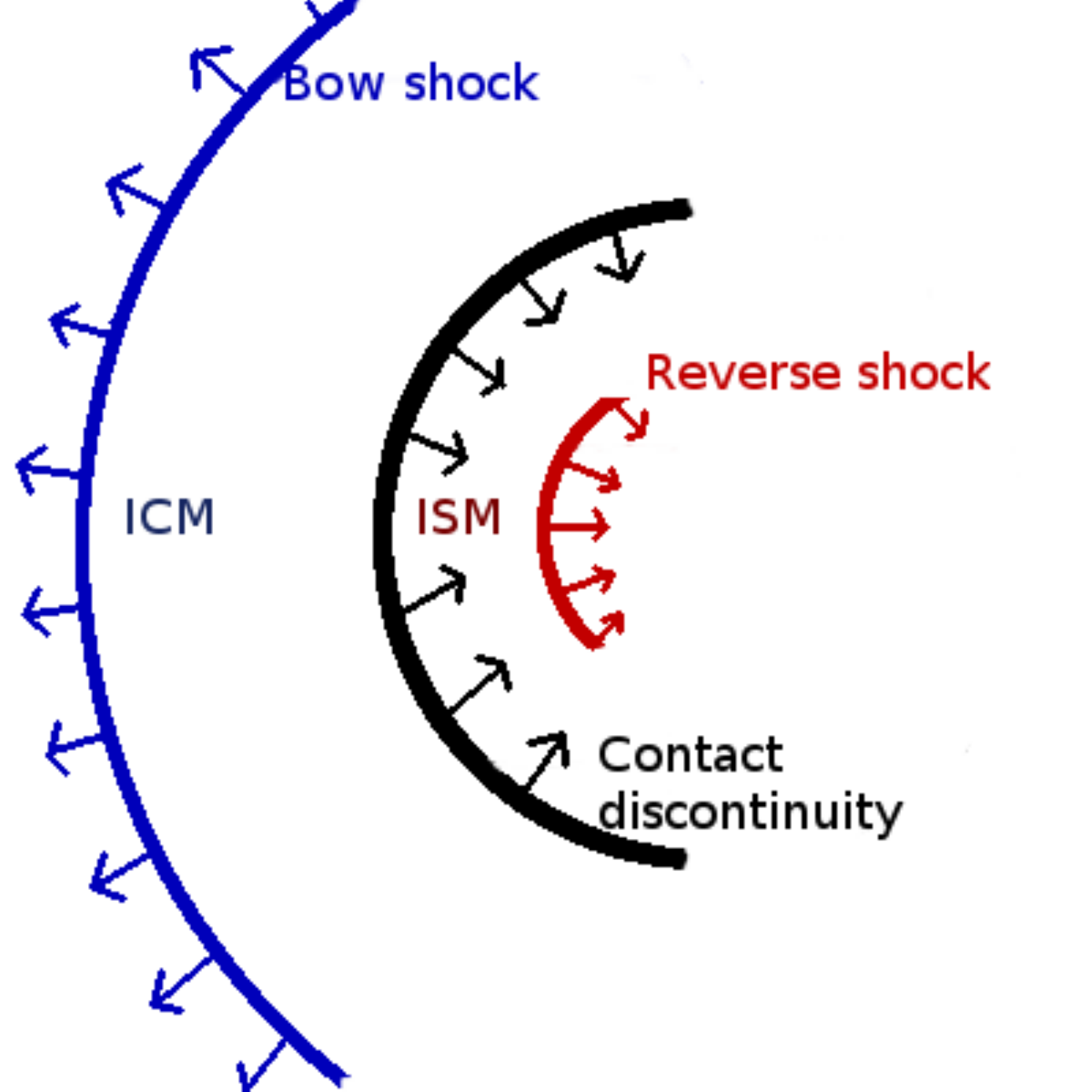}}
     \caption{Sketch illustrating the relative positions and directions of propagation of the bow shock, contact discontinuity at the ICM-ISM interface, and reverse shock.
      \label{fig:galaxyshock}} 
\end{figure}

\begin{figure*}[!htbp]    
       \begin{center}
{\includegraphics[width=\textwidth]{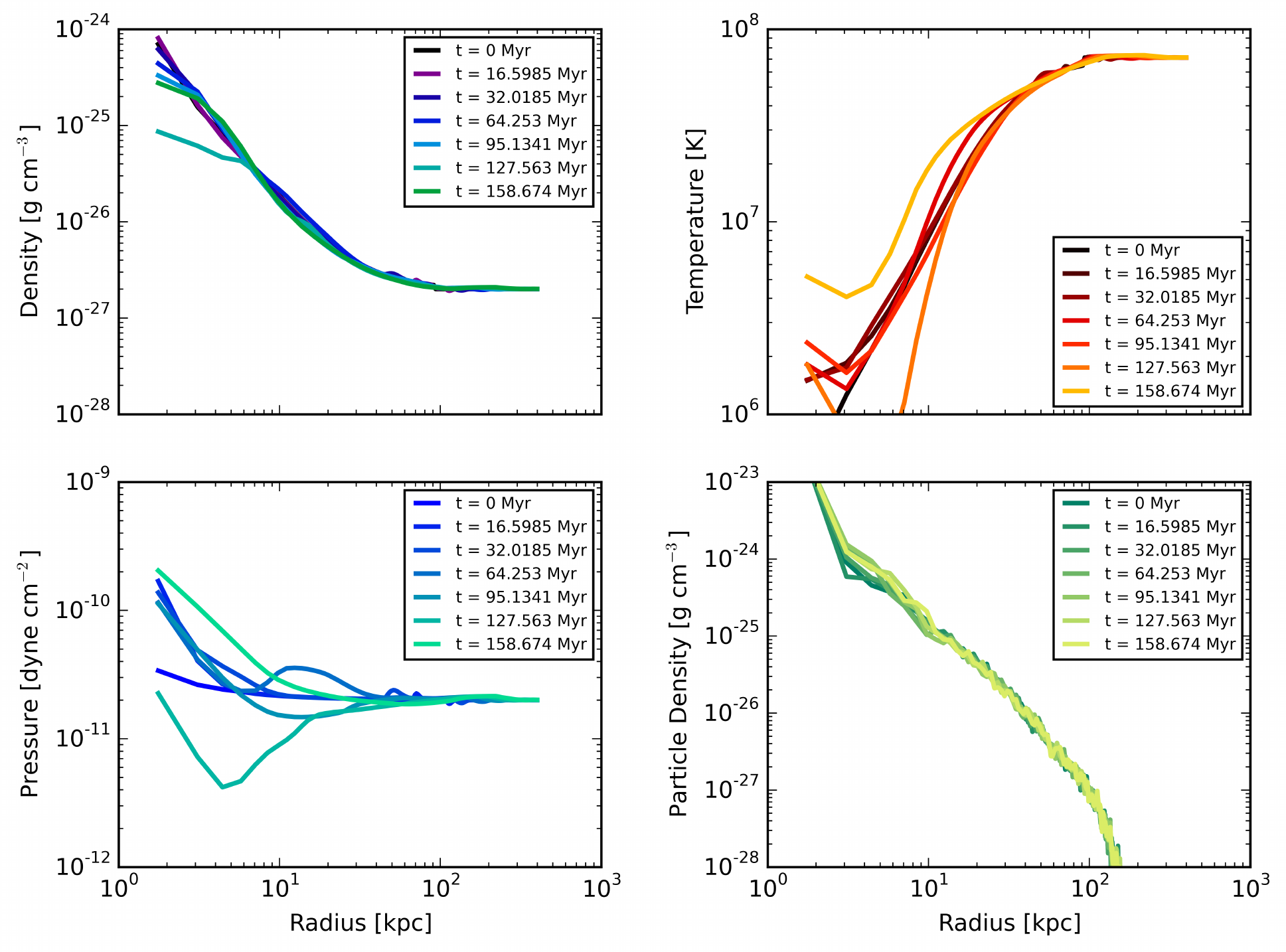}}
     \caption{Profiles in simulation with ram pressure but no thermal conduction.
      \label{fig:tprofwindnocond}} 
  \end{center}  
\end{figure*}

In this section, we discuss the stripping of a galaxy's hot ISM gas due to ram pressure.  We performed a base simulation run of the model galaxy in an ICM wind tunnel without thermal conduction. Under these simulated conditions, ram pressure due to the ICM wind is the only physical process by which the galaxy can lose its gas.
At $t = 0$ Myr, the density, temperature, and pressure of the ICM and ISM are continuous at the galaxy's virial radius, $R_{200}$. The galaxy and the ISM are at rest in the frame of the simulation box. The simulation box is three dimensional and cubic, with each side of length $2 \times 10^{24} $ cm, or $648$ kpc. The ICM wind flows into the box through the $-\mathbf{x}$ face. The other 5 faces of the box have outflow boundaries. The ICM wind has a velocity $\mathbf{v}_{\rm ICM} = v_{\rm ICM} \hat{x}$, with $v_{\rm ICM} = 610$ km s$^{-1}$. 

In more massive clusters, galaxies' velocities can be higher, $\ga 10^3$~km s$^{-1}$. Since ram pressure $P_{\rm ram} \propto \mathbf{v}_{\rm ICM}^2$, at higher velocities, and therefore in massive clusters, galaxies will stripped of their gas at a faster rate \citep[shown in][]{Vijayaraghavan15b}. Additionally, in realistic cluster environments, galaxies' orbital velocities will vary considerably unless they are on circular orbits, resulting in a range of ram pressure values encountered by galaxies. However, most of the galaxies' gas loss occurs during their first orbit within their host cluster ($t \sim 2 - 3$ Gyr), when galaxies' outer halos are stripped rapidly until their internal thermal pressure balances the sum of the ICM thermal pressure and ram pressure \citep{Vijayaraghavan15b}, followed by a slower disruption of the core. Therefore, the timescale for gas loss is determined by both the maximum ram pressure experienced by galaxies, as well as when galaxies encounter this ram pressure.

Figure~\ref{fig:slicenocond} shows slices of the ISM and ICM density, temperature, and $x$ component of the velocity in the $z = 0$ plane. Figure~\ref{fig:slicenocondlate} shows density slices at later times.
There is an initial contact discontinuity at the galaxy's virial radius on the side facing the incoming ICM wind. At this interface, as the ICM flows past the galaxy, shear instabilities leading to Kelvin-Helmholtz (KH) rolls are formed. A bow shock propagates out from the leading surface of the galaxy into the ICM (illustrated in Figure~\ref{fig:galaxyshock}). A reverse shock propagates inwards to the center of the galaxy, and penetrates the core of the galaxy at $t = 72$ Myr. The reverse shock is refracted around the core and converges behind the core (at $t = 88 - 95$ Myr). The density and pressure within the core drop significantly after the reverse shock has passed through the core ($t = 100 - 120$ Myr). Gas flows into the core to maintain hydrostatic equilibrium, and a second shock then propagates outward from where the initial reverse shock converged, through the core. As this shock passes through the core, the core gas is compressed resulting in increased core density and pressure. This compression delays the stripping of the core. The second shock propagates behind the initial bow shock; there is no subsequent reverse shock. The force due to ICM ram pressure pushes on the outer diffuse ISM and strips it away. The stripped ISM gas trails the galaxy initially in a hollow cylindrical tail.

After the lower density, outer ISM gas is stripped, the denser core ISM gas is stripped away at a  much slower rate (from $t \simeq 300 $ Myr to $t \simeq 1600$ Myr) and trails the galaxy in a single tail. This second narrow filled tail is also unstable to KH instabilities. The flow in these regions becomes turbulent and forms vortices, dissipating into the ICM within $t \simeq 1600$ Gyr. The ISM is completely stripped by $t \simeq 2400$ Gyr (seen in Figure~\ref{fig:slicenocondlate}).  Figure~\ref{fig:tprofwindnocond}  shows azimuthally averaged radial profiles of the gas density, temperature, pressure, and dark matter (particle) density from $t = 8 - 160$ Myr.  

\subsection{Thermal Conduction, No Ram Pressure}
\label{sec:results_cond}

\begin{figure*}[!htbp]
  \begin{center}  
    	{\includegraphics[width=\textwidth]{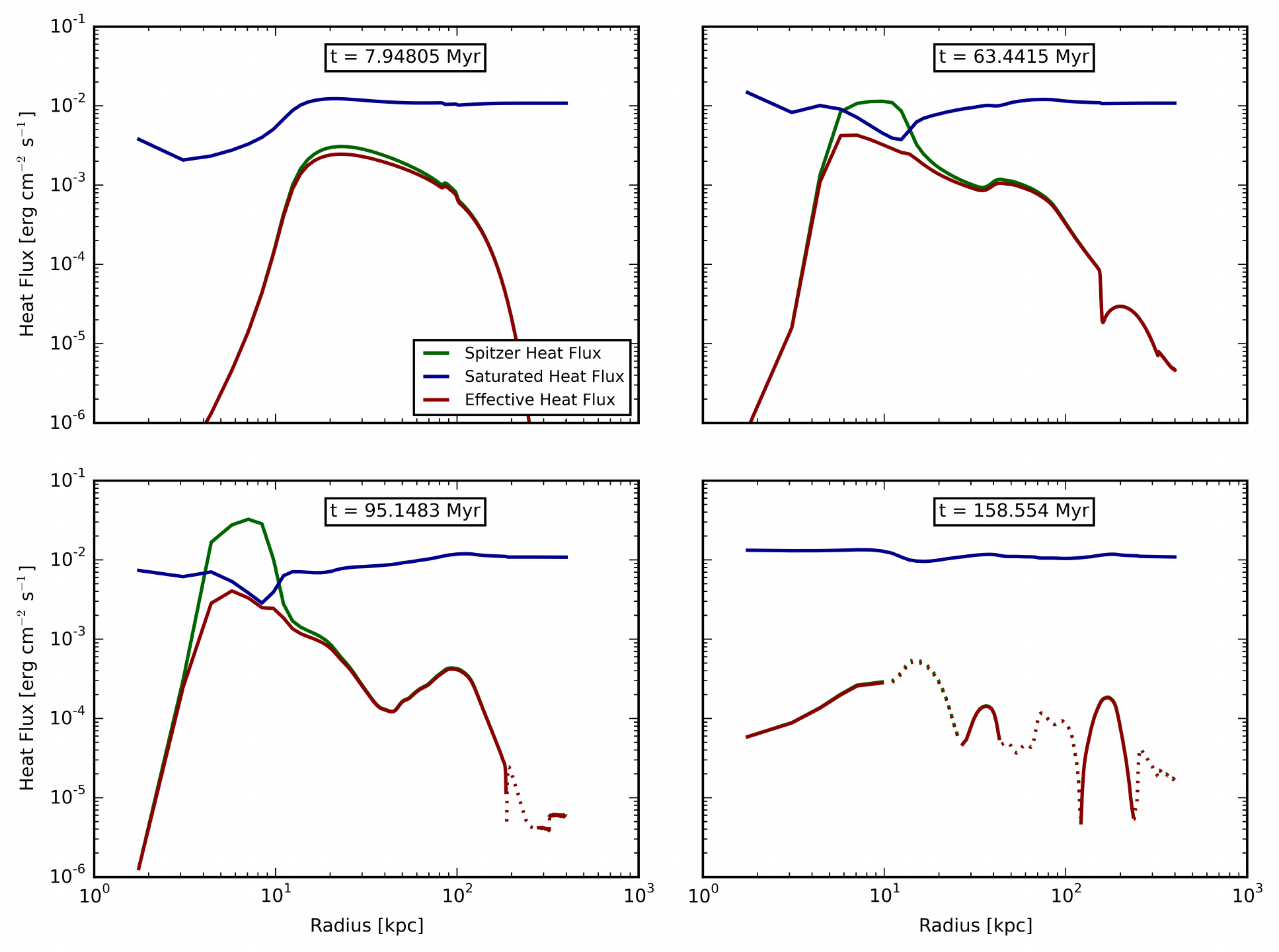}} 
     \caption{Profiles of the absolute values of the effective heat flux and expected theoretical values of saturated and classical Spitzer heat fluxes at different periods of the galaxy's evaporation. Solid lines correspond to a negative heat flux and dashed lines to a positive heat flux. \label{fig:heatfluxprofile}}
  \end{center}  
\end{figure*}

\begin{figure*}
  \begin{center}
	{\includegraphics[width=\textwidth]{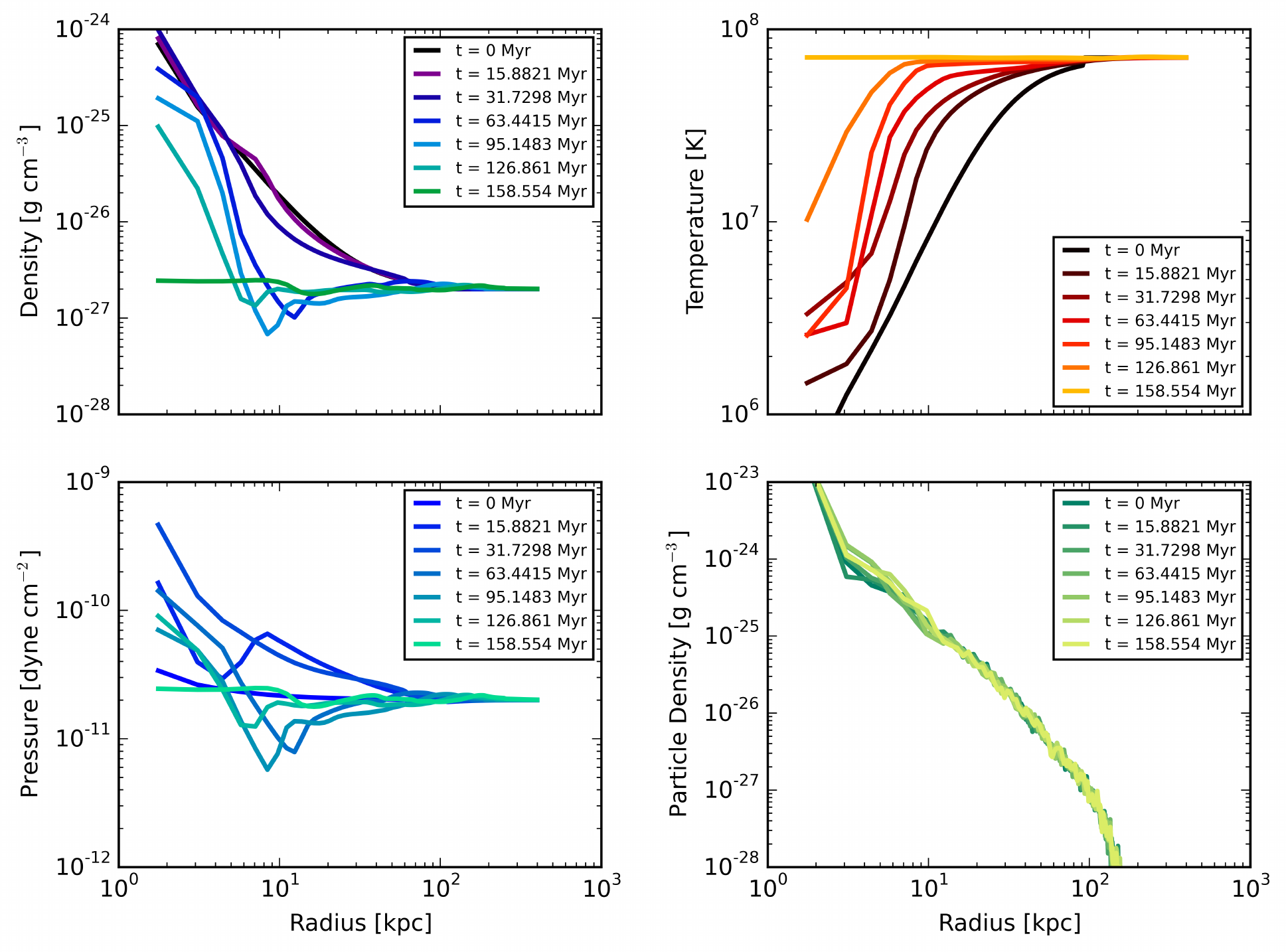}}
     \caption{Profiles in simulation of evaporation due to thermal conduction but no ram pressure. \label{fig:tprofhsecond}}
  \end{center}  

  \begin{center}
    {\includegraphics[width=\textwidth]{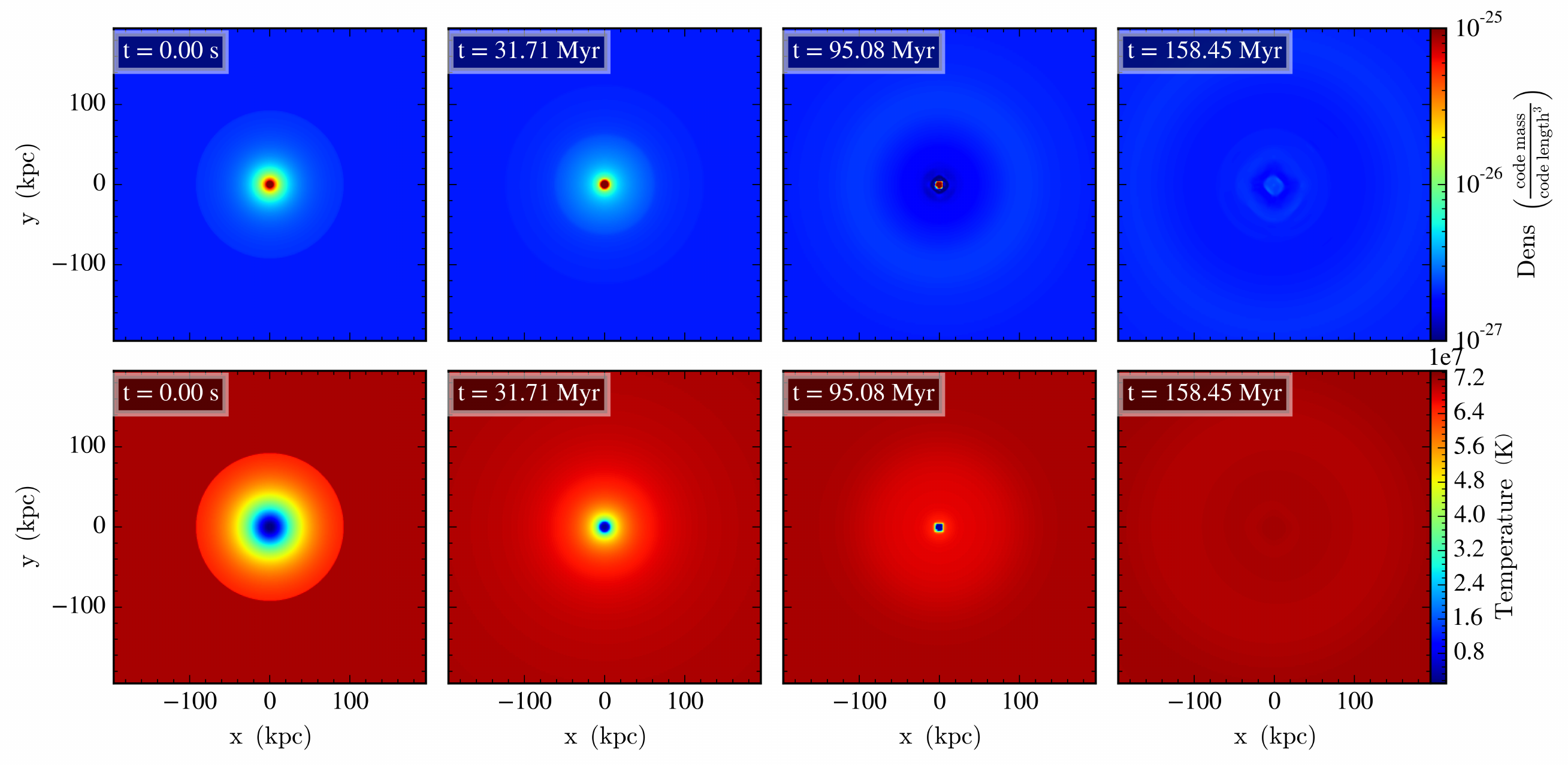}\label{fig:denstemphsecond1}}
     \caption{Slices of density and temperature in simulation of evaporation due to thermal conduction but no ram pressure. \label{fig:slicehsecond}}
  \end{center}  
\end{figure*}

In this section we discuss the effect of thermal conduction between the ICM and the ISM, conduction within the galaxy, and the subsequent evaporation of a galactic corona. We performed a simulation of the $2.8 \times 10^{11} \msun$ galaxy embedded in a static ICM, with $\mathbf{v}_{\rm ICM} = 0$ km s$^{-1}$. The other ICM parameters are as above. 

Figure~\ref{fig:heatfluxprofile} shows radial profiles of the absolute values of the effective heat flux, as well as the theoretically expected saturated heat flux and full Spitzer heat flux calculated from the density and temperature of the fluid at approximately $t = 8$ Myr, $t = 63$ Myr, $t = 95$ Myr, and $t = 159$ Myr. Dashed lines indicate regions where the heat flux is positive (i.e., a negative temperature gradient), while solid lines show the typical situation where the heat flux is negative (i.e., inwardly directed). Note that the heat flux is only positive in regions where it is comparatively weak, and where the temperature gradient is very close to zero (Fig.~\ref{fig:tprofhsecond}). The magnitude of the heat flux depends strongly on the temperature and temperature gradient as $Q \propto T_e^{5/2} \nabla T_e$. Initially, at $t = 8$ Myr, the effective heat flux is highest inside the galaxy, where there is a non-zero temperature gradient, and decreases dramatically outside the galaxy ($r \gtrsim 100 $ kpc) where the temperature profile is effectively flat. Within the galaxy, the effective heat flux depends sensitively on temperature, and is significantly higher outside the cool galaxy core than within the core, i.e., at $r \gtrsim 10$ kpc. 

Heat is transported from the ICM to the ISM and from the outer regions of the galaxy inwards. As the ISM is heated, it expands. Figure~\ref{fig:tprofhsecond} shows the evolution of density, temperature, and pressure profiles during the evaporation process. As heat is transported into the core, the core temperature increases, as do the density and pressure up to $t = 32$ Myr. From $t = 32 - 96$ Myr, the ISM density and pressure mostly decrease as the ISM expands, and the temperature increases up to the ICM temperature. After the initial expansion, gas flows back into the core briefly resulting in an increased core density and pressure between $t = 140 - 150$ Myr. This gas then expands and evaporates outwards; by $t = 160$ Myr, the density, temperature, and pressure profiles are fairly flat.

During this period of evaporation, the effective heat flux evolves considerably. At $t \simeq 55 - 135$ Myr, (Figure~\ref{fig:heatfluxprofile}) the core gas temperature has increased to the extent that the effective heat flux, now saturated, is significantly higher than at early times. Heat is transported to the core ISM gas from the ISM in the outskirts and the ICM. There are three distinct radial zones in the galaxy: an inner classical zone, where the gas is still cool enough that the classical Spitzer heat flux is lower than the  saturated flux ($r < 2 - 5$ kpc), an outer classical zone where the heat flux is primarily determined by the classical Spitzer heat flux since the temperature gradient is small ($r > 10 - 20$ kpc), and the intermediate saturated zone ($5 \lesssim r \lesssim 15$ kpc) where the temperature is sufficiently high and the temperature gradient is steep enough so the heat flux is saturated. \citet{Cowie77} describe these zones in their calculations of the evaporation of a cold gas cloud. For $t > 150 $ Myr, when sufficient heat has been transported from the ICM so that the galaxy's temperature profile is flat, the classical Spitzer heat flux is two orders of magnitude lower than the saturated heat flux.

The dark matter mass, which is an order of magnitude higher than the ISM mass, is mostly unaffected by the evaporation of the ICM, as seen in the particle density radial profile plot in Figure~\ref{fig:tprofhsecond}. The gravitational potential is primarily determined by the dark matter distribution, and therefore also does not change with the evaporation of gas. Therefore, to maintain hydrostatic equilibrium, the density and pressure gradients, while small, are non-zero. For $\Delta P = P(R_{200}) - P(0)$  and $\Delta \rho_{\rm gas} = \rho_{\rm gas}(R_{200}) - \rho_{\rm gas}(0)$, with $\Delta P \ll P $ and $\Delta \rho_{\rm gas} \ll \rho_{\rm gas}$, the equation of hydrostatic equilibrium can be written as:
\begin{equation}
\frac{\Delta P}{R_{200}} \simeq \rho_{\rm gas} \frac{GM}{R_{200}^2}.
\end{equation}
Additionally, the ideal gas law gives us:
\begin{equation}
\Delta P = \frac{k_B}{\mu m_p} \Delta \rho_{\rm gas} T, 
\end{equation}
where we have dropped $\Delta T$ since $\Delta T / T \ll \Delta P / P $ and
$\Delta T / T \ll \Delta \rho_{\rm gas} / \rho_{\rm gas} $.
Therefore, we have:
\begin{equation}
\frac{\Delta P}{P} \simeq \frac{\Delta \rho_{\rm gas}}{\rho_{\rm gas}}.
\end{equation}
\newline \\

At $t = 400$ Myr, the density, pressure, and temperature profiles of the simulated galaxy correspond to $\dfrac{\Delta P}{P} \simeq \dfrac{\Delta \rho_{\rm gas}}{\rho_{\rm gas}} \simeq 0.02$, while $\dfrac{\Delta T}{T} \simeq 2 \times 10^{-4}$, consistent with our earlier assumptions of $\Delta T / T \ll \Delta P / P $ and $\Delta T / T \ll \Delta \rho_{\rm gas} / \rho_{\rm gas} $.

The heating, expansion, and evaporation of the ISM are seen in the density and temperature slices of Figure~\ref{fig:slicehsecond}. We see the expanding shell of gas corresponding to the evaporating ISM at $t = 96$ Myr and $t = 160$ Myr. The galaxy's dense, cool core also visibly evaporates by $t = 160$ Myr.

\citet{Cowie77} determined the mass loss rate of a cool, spherical gas cloud of radius $R$, number density $\bar{n}_c$ with $T \sim 0$ at $r = R$, embedded in a hot medium with $T = T_f$ at $r \gg R$. Assuming classical evaporation with Spitzer conductivity, they show that the evaporation timescale is:
\begin{equation} \label{eqn:tevap}
t_{\rm evap}
= 3.3 \times 10^{20} \bar{n}_c R^2 T_f^{-5/2} \frac{\ln \Lambda}{30} \rm{Myr}.
\end{equation}
For a gas cloud with the properties of the ISM in our simulated galaxy, where $R =  R_{200}$ (in kpc) and $\bar{n_c} = M_{\rm gas}/(\frac{4}{3} \pi R_{200}^3 \times \mu m_p), \, t_{\rm evap} = 216$ Myr, comparable to the actual evaporation timescale of $t = 160$ Myr. This evaporation timescale for gas flowing out from a cool cloud assumes steady state and does not include the effect of gravity. These assumptions do not hold for a galaxy with a potential gradient primarily determined by dark matter, which is not affected by thermal conduction. In our simulations, the galaxy's gravitational potential and gas are initially in hydrostatic equilibrium. As heat diffuses into the ISM, the ISM becomes isothermal and its density, pressure, and temperature profiles flatten, with a small gradient, but the ISM never evaporates completely. This simulation does not include the effect of ram pressure or tidal forces from the cluster's gravitational potential, which are ultimately responsible for removing all of the ISM gas. However, we find that if thermal conduction with saturation is in fact effective, galaxies should not have prominent dense coronae.

\subsection{Gas Loss Due to Thermal Conduction and Ram Pressure}
\label{sec:results_windcond}

\begin{figure*}
  \begin{center}
{\includegraphics[width=\textwidth]{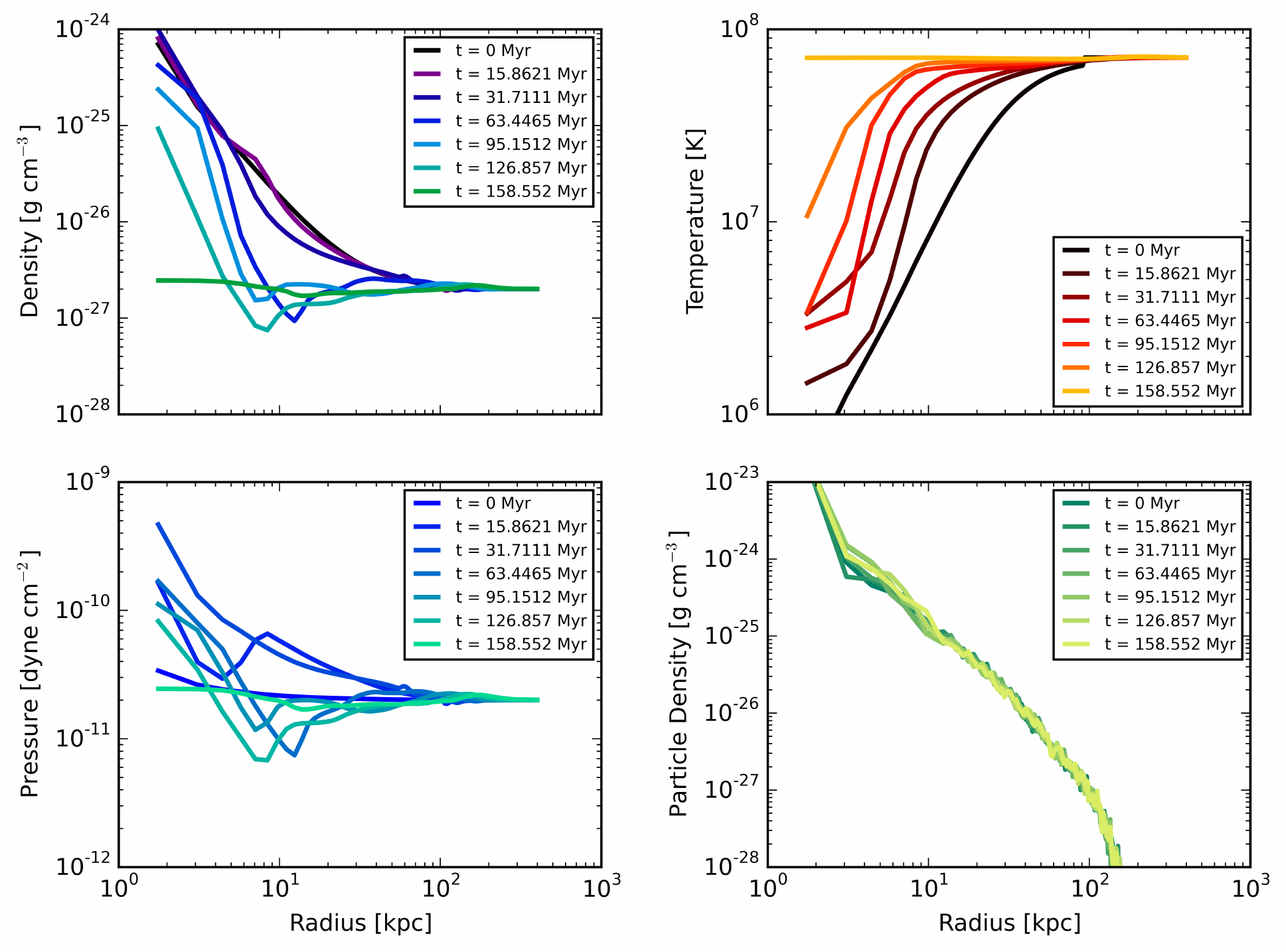}}
     \caption{Profiles in simulation of evaporation due to thermal conduction with ram pressure.\label{fig:tprofhsewindcond}}
  \end{center}  

  \begin{center}
	 {\includegraphics[width=\textwidth]{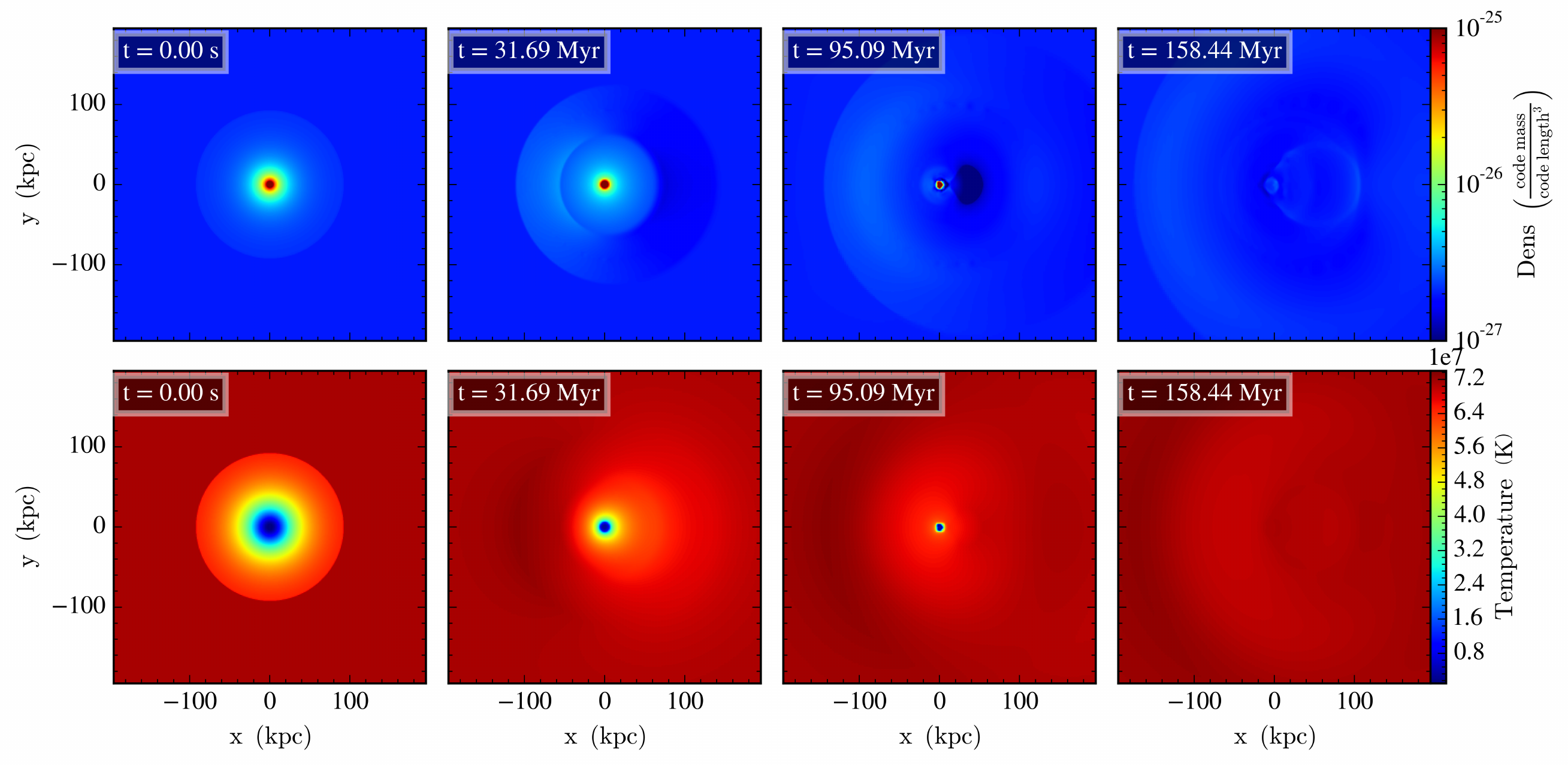}\label{fig:denstemphsewindcond1}}
     \caption{Slices of density and temperature in simulation of evaporation due to thermal conduction with ram pressure. \label{fig:slicehsewindcond}}
  \end{center}  
\end{figure*}

We see in \S~\ref{sec:results_nocond} that ram pressure alone removes the ISM by $t = 2$ Gyr, while isotropic thermal conduction results in the evaporation of the ISM by $t = 160$ Myr, as seen in \S~\ref{sec:results_cond}. Therefore, evaporation due to thermal conduction is much more rapid than ram pressure stripping in removing gas. The presence of flowing ICM does however qualitatively affect the process of evaporation due to thermal conduction. Thermal conduction suppresses the formation of shear instabilities at the cool ISM -- hot ICM interface, as heat is transported from the ICM to the outer edge of the ISM before KH instabilities develop.  Thermal conduction broadens the transition region between the ISM and ICM, which smears out the velocity shear.  This makes the ICM--ISM interface less Kelvin-Helmholtz unstable, particularly for the fastest growing, short wavelength modes \citep{Cha61,RKF+13}.

\begin{figure*}[!htbp]
  \begin{center}
{\includegraphics[width=\textwidth]{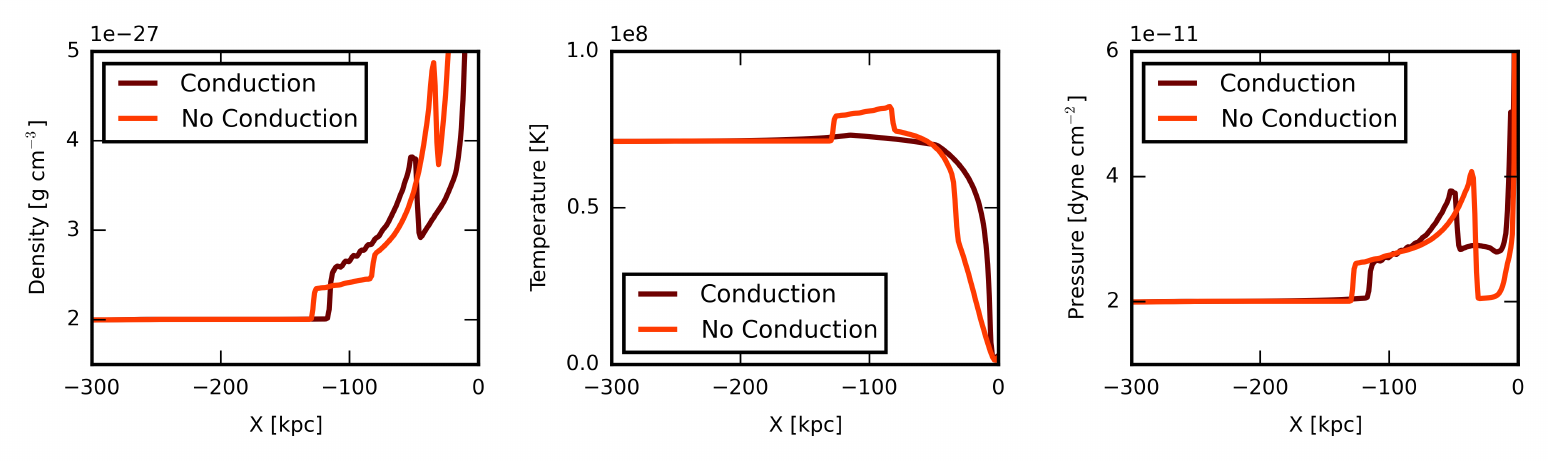}}
     \caption{Density, temperature, and pressure along the $x$-axis showing the presence of the shock and contact discontinuity in simulations with ram pressure, in the presence or absence of thermal conduction at $t = 40$ Myr.\label{fig:shockjump}}
  \end{center}  
\end{figure*}

There is also a bow shock and a contact discontinuity, as in the simulation without thermal conduction. However, the nature of the bow shock is different in the presence of thermal conduction.
A a result of rapid, efficient, heat transport, the shock is isothermal. These are seen in the density, temperature, and pressure jumps illustrated for both cases at $t = 40$ Myr in Figure~\ref{fig:shockjump}. The shock front is visible in the density and pressure profiles at $x = -130$ kpc in the case with no conduction, and at $x = -115$ kpc in the case with thermal conduction. We see in Figure~\ref{fig:shockjump} that there is no temperature jump at the shock front in the presence of thermal conduction. Additionally, the contact discontinuity at $R_{200} = 90$ kpc at the ICM-ISM interface is clearly visible as a density jump with no corresponding pressure jump. The shock front in the case with conduction trails the shock in the absence of conduction, therefore the isothermal shock is slower than the adiabatic shock. The Mach numbers of the shocks in both cases are however comparable. The Mach number of the adiabatic shock $M_{\rm adiabatic} = v_{\rm shock, adiabatic} / c_{\rm adiabatic}$, where the sound speed $c_{\rm adiabatic}  = \sqrt{\gamma P / \rho}$, is $M_{\rm adiabatic} =  1.136$. The Mach number of the isothermal shock  $M_{\rm isothermal} = v_{\rm shock, isothermal} / c_{\rm isothermal}$, where the sound speed $c_{\rm isothermal}  = \sqrt{P / \rho}$, is $M_{\rm isothermal} =  1.145$.

The bow shock expands outwards and dissipates, while the galaxy's core gas evaporates due to thermal conduction. The core evaporates before ram pressure strips it away. The reverse shock passes through the evaporating, almost isothermal core, and converges behind the core. This reverse shock then expands outwards in a thin shell, eventually dissipating by $t \simeq 200$ Myr. The galaxy's core evaporates before the expanding reverse shock passes through it. Compared to a gas loss timescale of $t = 2000$ Myr due to ram pressure stripping alone, in the presence of thermal conduction, the galaxy's ISM has evaporated by $t = 200$ Myr; the remnant gas in the galaxy's potential well is isothermal with the ICM, with a small gradient in the density and pressure profiles. This gas, as in \S~\ref{sec:results_cond}, never has a completely flat density and temperature profile as long as hydrostatic equilibrium is effective. Of course, in real clusters tidal and other gravitational effects will flatten the potential and remove gas in addition to any evaporation due to thermal conduction. 

The evolution of density, pressure, temperature, and dark matter density profiles of this galaxy are shown in  Figure~\ref{fig:tprofhsewindcond}. These azimuthally averaged profiles are nearly identical of the evolution of fluid variable profiles in Figure~\ref{fig:tprofhsecond}, showing that the presence of an ICM wind and strong ram pressure is not as effective at removing gas as isotropic, saturated thermal conduction.

\subsection{Varying Galaxy Mass and ICM Density}
\label{sec:results_massdensity}

To test the qualitative and quantitative variations in the evaporation of galaxies due to thermal conduction, we performed a simulation of a galaxy 10 times more massive than our base model, and of our  standard lower mass galaxy in a low density, low pressure ICM of the same temperature. These are lower resolution simulations, corresponding to a maximum spatial resolution of 2.53 kpc, a factor of two higher than the 1.26 kpc resolution in the previously described simulations. This resolution is still significantly smaller than the ICM mean free path, and the overall results are unaffected.

\subsubsection{High Mass Galaxy}
\label{sec:results_massdensity_highmass}

\begin{figure*}[!htbp]
  \begin{center}
 {\includegraphics[width=\textwidth]{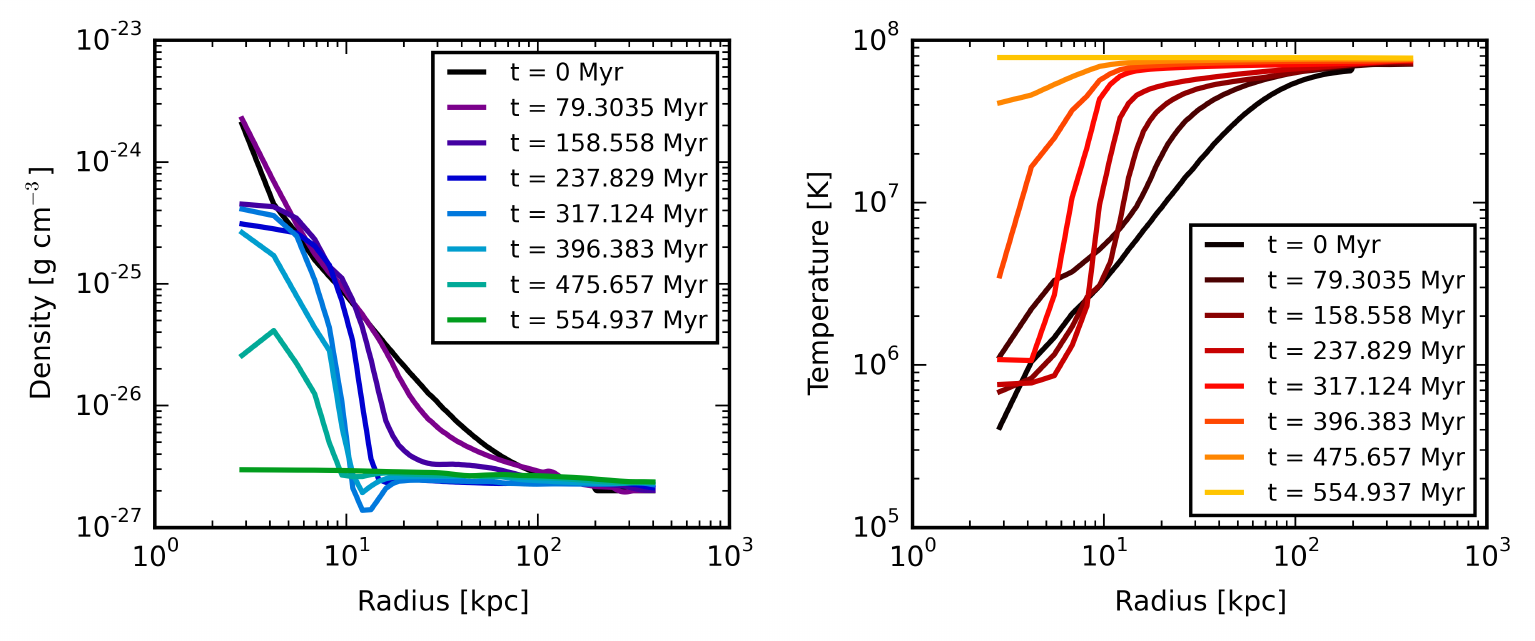}}
     \caption{Profiles of gas density and temperature in a simulation of evaporation due to thermal conduction with ram pressure for a more massive galaxy.
     \label{fig:tprofmassivegalaxycond}}
  \end{center}  

  \begin{center}
{\includegraphics[width=\textwidth]{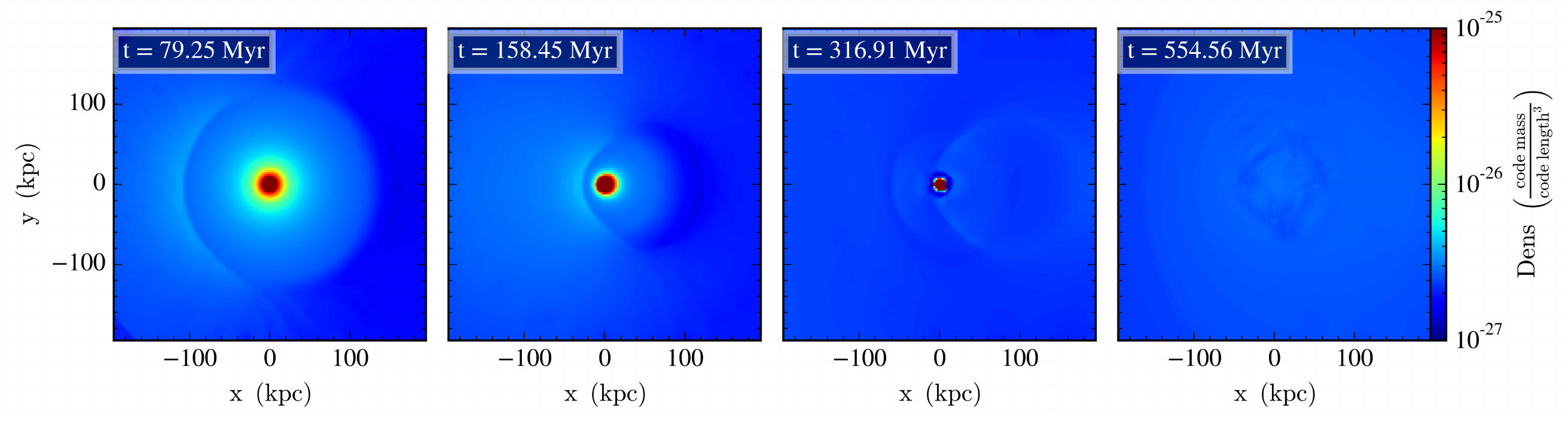}\label{fig:denstempmassivegalaxycond1}}
     \caption{Slices of the gas density in a simulation of evaporation due to thermal conduction with ram pressure for a $2.8 \times 10^{12} \msun$ massive galaxy.
      \label{fig:slicemassivegalaxycond}}
  \end{center}  
\end{figure*}

We simulated the evaporation of a  $2.8 \times 10^{12} \msun$ galaxy in an ICM wind tunnel identical to the previous simulations. This galaxy has a virial radius $R_{200} = 190 \, \rm{kpc}$. To adequately capture the expanding bow shock, we use a larger simulation box of side $1296$ kpc in which all of the spatial dimensions are twice as large as in the earlier simulations. Although the average ISM density as well as the ICM density, pressure, and temperature are identical to the previous runs, the larger galaxy size affects the evaporation timescale.
The overall evaporation time should scale, approximately, as $t_{\rm evap} \propto M_{\rm gas} / R_{200}$ (equation ~\ref{eqn:tevap}). Compared to the standard galaxy model, this galaxy's mass is 10 times higher, and its radius is twice as large, therefore the evaporation time should approximately five times longer. The total evaporation time for this galaxy is $t \simeq 555$ Myr, while the evaporation time for the fiducial galaxy is $t \simeq 160$ Myr.  This timescale is still significantly lower than the ram pressure stripping timescale. The evaporation time is $\sim 3.5 \times $ higher for the massive galaxy. This is consistent with the analytic idealized predictions of Equation~\ref{eqn:tevap}, where at a given density and ICM temperature, $t_{\rm evap} \propto R^2$. The virial radii of the fiducial galaxy and massive galaxy are $R_{200} = 92 \, \rm{kpc}$ and $R_{200} = 190 \, \rm{kpc}$ respectively, therefore the evaporation time will scale by a factor of $\sim 2^2 = 4$.

The evolution of the ISM is seen in azimuthally averaged profiles of density and temperature and slices of density in Figures~\ref{fig:tprofmassivegalaxycond} and ~\ref{fig:slicemassivegalaxycond}. The timesteps at which density and temperature values are shown here are different from those in the previous runs, since the overall evaporation time is higher. Qualitatively, the evolution of the ISM is similar to the galaxy in \S~\ref{sec:results_windcond}. There is an initial contact discontinuity at the ISM-ICM interface on the side of the galaxy facing the ICM wind. A bow shock propagates outward into the ICM, while a reverse shock propagates into the galaxy. The reverse shock converges behind the core, then expands outward.
Through this process, the ISM is also heated due to thermal conduction with the ICM, and continues to expand. In the simulations in \S~\ref{sec:results_cond} and \S~\ref{sec:results_windcond}, the galaxy's core evaporates before the expanding reverse shock, after converging behind the galaxy, passes through the core. The massive galaxy's core, however, does not evaporate until after the shock passes. The expanding reverse shock passes through the core  at $t \simeq 325$ Myr. The core expands and evaporates at $t = 515$ Myr, forming a low density cavity; this evacuated zone is soon filled with inflowing ISM gas and by $t = 555$ Myr the ISM has flat density, pressure, and temperature profiles. As in the earlier runs, the dark matter density profile is largely unaffected.

\subsubsection{Low Density ICM}
\label{sec:results_massdensity_lowdensity}

\begin{figure*}[!htbp]
  \begin{center}
  {\includegraphics[width=\textwidth]{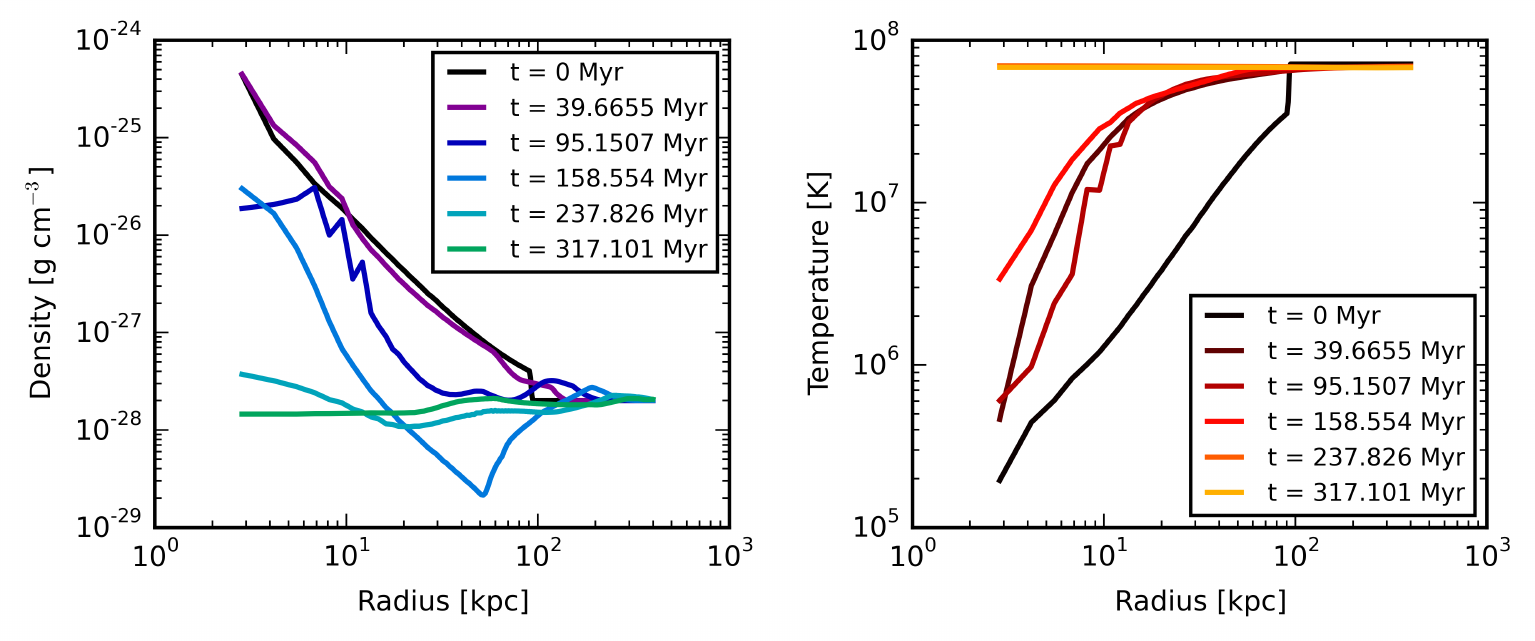}}
     \caption{Profiles of gas density and temperature in a simulation of evaporation due to thermal conduction with ram pressure for a lower ICM density.
      \label{fig:tproflowdensityICMcond}}
  \end{center}  

  \begin{center}
	{\includegraphics[width=\textwidth]{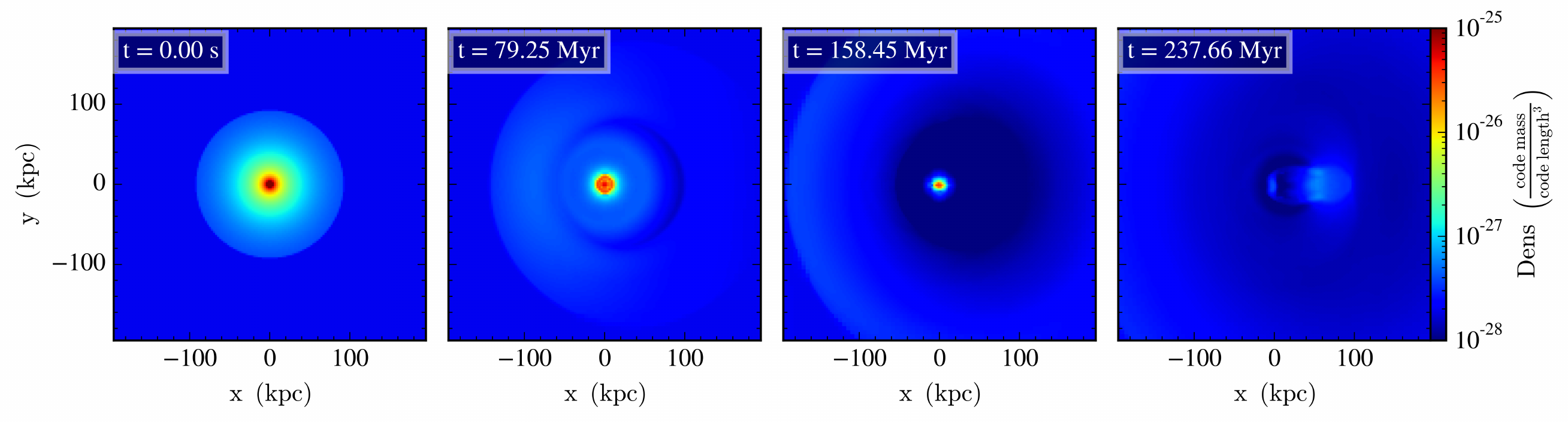}\label{fig:denstemplowdensityICMcond1}}
     \caption{Slices of gas density in a simulation of evaporation due to thermal conduction with ram pressure for a lower ICM density. \label{fig:slicelowdensityICMcond}}
  \end{center}  
\end{figure*}

\begin{figure*}
  \begin{center}
 	{\includegraphics[width=\textwidth]{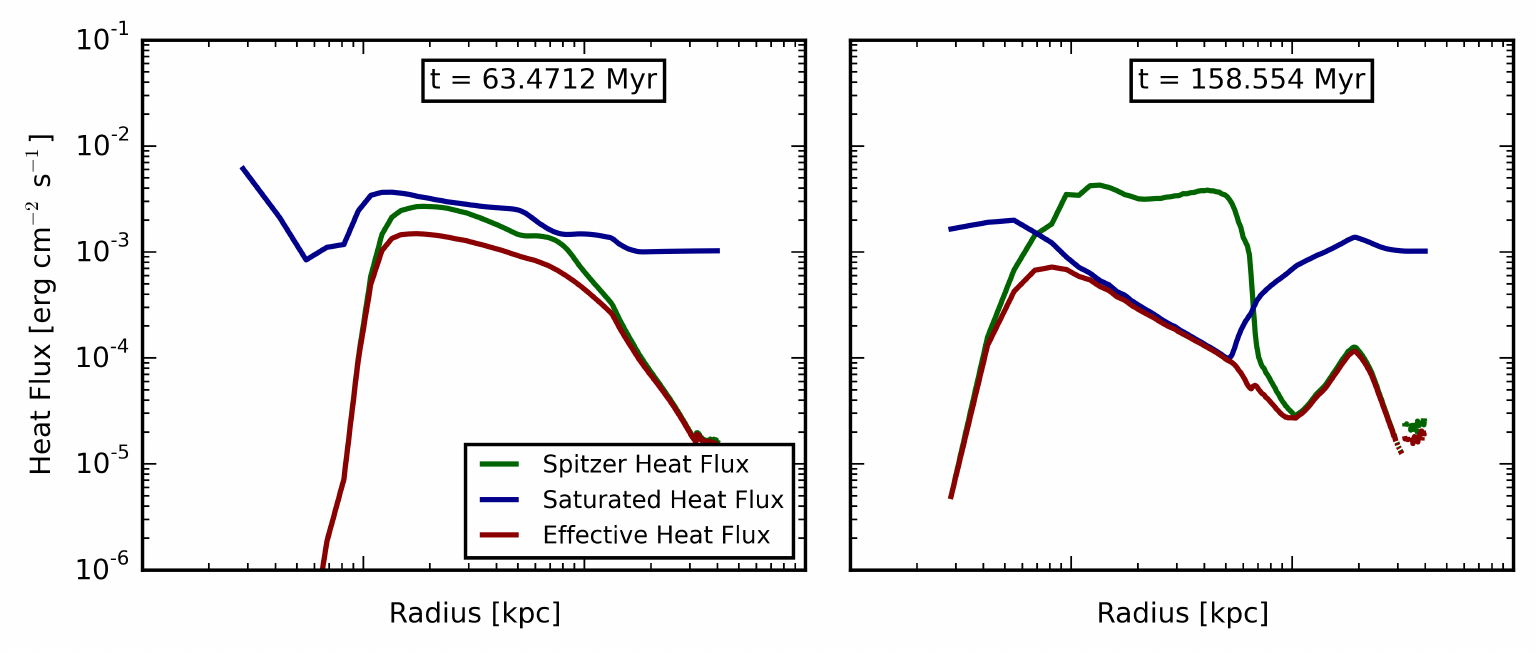}}  
     \caption{Profiles of the effective heat flux and expected theoretical values of saturated and classical Spitzer heat fluxes at different periods of the galaxy's evaporation in the low density ICM. \label{fig:heatfluxprofilelowdensity}}
  \end{center}  
\end{figure*}

We performed a simulation of the $2.8 \times 10^{11} \msun$ galaxy in an ICM wind tunnel with $\rho_{\rm ICM} = 2 \times 10^{-28}$ g cm$^{-3}$, $P_{\rm ICM} = 2 \times 10^{-12}$ dyne cm$^{-2}$, and $T_{\rm ICM}$ identical to the previous simulations at $7.14 \times 10^7$ K. The ICM wind speed is $v_{\rm ICM} = 610$ km s$^{-1}$, and this simulation includes saturated thermal conduction. 
Since the ISM's density profile is initialized to be continuous with the ICM density, the density in the galaxy's outskirts is lower than in the previous simulation; the pressure profile is also modified to maintain hydrostatic equilibrium. Subsequently, the temperature in the galaxy's outskirts is lower than in the ISM in our earlier simulations, seen when comparing the $t = 0$ Myr temperature profiles in Figure~\ref{fig:tprofhsecond}.

The cooler ISM and lower density lead to lower heat fluxes due to thermal conduction. This is seen in Figure~\ref{fig:heatfluxprofilelowdensity}, which shows radial profiles of the effective, classical, and saturated heat fluxes at $t = 64$ and $t = 160$ Myr. The typical effective heat fluxes at these times are $|\mathbf{Q_{\rm eff}}| \simeq 10^{-4} - 10^{-3}$ erg cm$^{-2}$ s$^{-1}$, while in the high density ICM,  $|\mathbf{Q_{\rm eff}}| \simeq 10^{-3} - 10^{-2}$ erg cm$^{-2}$ s$^{-1}$ at $t = 63$ Myr. At $t = 160$ Myr the ISM has evaporated in the high density ICM, and there is no temperature gradient.   Additionally, since the ICM wind speed is unchanged while 
the ICM density is lower by an order of magnitude, the ram pressure is also reduced by an order of magnitude. The net result is that at lower heat fluxes, the rate of evaporation of the galaxy is much lower; the time taken for the galaxy to evaporate is $t \simeq 350$ Myr. Qualitatively, the evolution of the ISM (Figure~\ref{fig:slicelowdensityICMcond}) is similar to the high density ICM case, seen in Figure~\ref{fig:slicehsewindcond}: an initial shock wave is formed at the ICM-ISM interface, propagates out into the ICM. The ISM expands as it is heated, and the density outside the core drops rapidly. The galaxy's core evaporates by $t = 250$ Myr. The ICM reverse shock converges behind the galaxy's core, then expands and dissipates. The time evolution of the density profile is also seen in Figure~\ref{fig:tproflowdensityICMcond}.

\section{Discussion}
\label{sec:discussion}

Galaxies' hot coronae evaporate on $\sim 10^2$ Myr timescales in the presence of isotropic thermal conduction including saturation in the intracluster medium. Specifically, in the simulations here, evaporation times  are $\sim 160$ Myr for the $2.8 \times 10^{11} \msun$ galaxy, and $\sim 550$ Myr for the $2.8 \times 10^{12} \msun$ galaxy. Ram pressure strips gas on much longer timescales. The gas loss time due to ram pressure stripping alone, in an identical ICM, is $\sim 2.5$ Gyr for the $2.8 \times 10^{11} \msun$ galaxy; the simulated galaxies evaporate due to conduction alone well before ram pressure can be effective. Qualitatively, by comparing the effects of evaporation due to conduction in \S~\ref{sec:results_cond} and \S~\ref{sec:results_windcond}, we see that the additional effect of ICM ram pressure is to form an isothermal bow shock as well as a reverse shock; evaporation rates are nearly identical in both cases. In the absence of ICM wind and ram pressure, there is no bow shock or reverse shock. Even when the conductive heat flux is much reduced, as in the lower density ICM, the complete evaporation time is $\sim 300$ Myr for the  $2.8 \times 10^{11} \msun$ galaxy. Efficient thermal conduction therefore results in the rapid evaporation of galaxies' hot X-ray emitting coronae, including their cool dense cores.

Galaxies moving relative to the ICM at higher velocities will experience stronger ram pressure as $P_{\rm ram} \propto \mathbf{v}_{\rm ICM}^2$. However, the thermal evaporation time scales are more than an order of magnitude shorter than the stripping time scales in our simulations. Thus, speeds of greater than 2000 km s$^{-1}$ would probably be required for ram pressure stripping to become competitive with thermal evaporation, all other things being equal. Higher galaxy speeds are expected in more massive clusters, but these also have higher gas temperatures, with $k T \propto \mathbf{v}_{\rm ICM}$. Since the thermal conduction and evaporation time scales decrease much faster than linearly with $T$ (e.g., equation~\ref{eqn:tevap}), the dominance of thermal evaporation is actually expected to be stronger in clusters with higher galaxy velocities.

In addition to the evaporation of coronae and central cores, thermal conduction  suppresses shear instabilities at the ISM-ICM interface, and prevents the formation of stripped tails. In the absence of conduction, ram pressure pushes on the ISM in the direction of the ICM wind, stripping the ISM away from the galaxy. Stripped ISM gas trails the galaxy, forming Kelvin-Helmholtz instabilities at the ICM interface. The galaxy's core ISM gas is subject to ram pressure once the outer ISM has been peeled away, and gradually dissipates. In our simulated galaxy, the formation of the initial tail takes $t \sim 300 - 400$ Myr (Figure~\ref{fig:slicenocondlate}). The total evaporation timescale for the same galaxy is $t = 160$ Myr, therefore stripped galaxies will not be able to form tails if thermal conduction is efficient. Even in the lower density ICM simulation, where thermal conduction is slower, there is no stripped tail. The complete suppression of the formation of hot X-ray emitting tails is in direct contradiction with observed X-ray tails in cluster galaxies \citep[e.g.,][]{Forman79,Irwin96,Sun05b,Machacek06,Randall08,Kim08,Sun10,Kraft11,Zhang13}.

In addition to the absence of stripped tails, thermal conduction results in the rapid removal of galaxies' central coronae. The evaporation and absence of coronae is also in tension with observations of compact X-ray emitting coronae in a large fraction of group galaxies ($\sim 80\%$ of $L_K > L_*$ galaxies; \citealt{Jeltema08}) and cluster galaxies ($\sim 60\%$ of $L_K > 2 L_*$ galaxies; \citealt{Sun07}). The frequency of these observations suggests that observed coronae are long-lived, i.e., they survive stripping and evaporation for many Gyr. Given the prevalence of stripped tails and X-ray emitting coronae, isotropic thermal conduction is therefore likely ineffective in ICM-ISM interactions in real clusters. The temperature and density structure of the ICM and hot ISM result in temperature gradients and electron mean free paths that favor efficient thermal conduction.  Evaporation must therefore be contained, or any gas loss must be rapidly and continually replenished given the long lifetimes of these coronae. 

Comparing the ram pressure stripping time ($\sim 2$ Gyr) with the evaporation time ($\sim 150$ Myr) for our simulated galaxy, we estimate that thermal conduction must be suppressed by at least a factor of $10 - 20$ to explain the observed X-ray tails in clusters, and considerably longer to sustain observed X-ray coronae for several Gyr. \citet{Vikhlinin01} estimate that conductivity must be suppressed by a factor of $\sim 100 \times$, for the cooling rate calculated from the X-ray emissivity for the central coronae in their paper. A more extensive review and calculations by \citet{Sun07} for satellites galaxies' coronae also show that conduction must be suppressed by a factor of $\sim 20 - 100$ in observed coronae. For the survival of observed cold fronts in clusters, calculations based on mean free path arguments, reviewed by \citet{Markevitch07}, require suppression factors of at least 3-4.

The presence of magnetic fields and the subsequent anisotropic thermal conduction along magnetic field lines (as opposed to isotropic thermal conduction) dramatically reduces evaporation rates; overall ISM evaporation timescales are comparable to ram pressure stripping timescales. We explore the effects of anisotropic thermal conduction in detail in Paper II. To understand the survival of coronae around central galaxies in the Coma cluster, \citet{Vikhlinin01} hypothesize that radiative cooling in principle can offset evaporation, if the rate at which gas cools is equal to the heating rate. The general applicability of this process is debatable since it requires identical heating and cooling rates to prevent either process completely destroying central coronae.

In addition to the evaporation of galactic coronae, thermal conduction has other effects on the evolution of the ICM. \citet{Jubelgas04} and \citet{Dolag04} find using SPH cosmological simulations of cluster formation that thermal conduction results in flat temperature profiles in the cores of simulated clusters and a smoother temperature distribution in the ICM. \citet{Smith13} find using grid-based numerical simulations that isotropic conduction results in isothermal cluster cores, but not in a significant reduction in temperature inhomogeneity. These simulations did not explicitly include magnetic fields; we discuss the additional effects of magnetic fields and anisotropic thermal conduction on the ICM in Paper II. 

\section{Conclusions}
\label{sec:conclusions}
We performed simulations of the loss of gas from galaxies moving relative to the ICM ($\rho_{\rm ICM} = 2 \times 10^{-27}$ g cm$^{-3}$, $T_{\rm ICM} = 7.14 \times 10^7$ K, $v_{\rm galaxy, ICM} = 610$ km s$^{-1}$) due to ram pressure and thermal conduction. The primary objective of these simulations was to quantify the effect of isotropic saturated thermal conduction on the evaporation of cluster galaxies' gas. Using simulations without conduction, we characterized the effects of ram pressure alone: a bow shock and reverse shock are initially driven from the ISM-ICM interface, the reverse shock converges behind the core of the galaxy before expanding outwards, Kelvin-Helmholtz instabilities are formed at this interface as the ICM flows past the ICM, the force due to ram pressure gradually unbinds the ISM from the ICM, the ISM trails the galaxy in a tail of gas which dissipates in the ISM over a total timescale of $t = 2-3$ Gyr for a $2.8 \times 10^{11} \msun$ galaxy.

Thermal conduction, in the absence of ram pressure, results in the transport of heat from the ICM to the ISM; as a result the outer ISM initially expands and evaporates, with an increase in temperature and decrease in density and pressure. The cooler, denser core is then heated; eventually, the ISM and the ICM are isothermal. The classical heat flux $Q \propto T_e^{5/2} \nabla T$, while the saturated heat flux $Q_{\rm sat} \propto T^{3/2}$ and is independent of the temperature gradient. As the ISM evaporates, thermal conduction is briefly saturated within the galaxy when the outer ISM is heated to high temperatures, resulting in a steep temperature gradient within the ISM. During this period, there is an inner classical zone of low temperature gas, an intermediate zone of hot, high temperature gradient gas, and an outer classical zone of hot, low temperature gradient gas. The background galaxy potential is unaffected in our simulations, particularly since we do not include the cluster's gravitational potential, therefore hydrostatic equilibrium forces a small temperature, density, and pressure gradient. The evaporation time for the hot ISM is $t = 160$ Myr for a $2.8 \times 10^{11} \msun$ galaxy. 

In the presence of ram pressure and isotropic thermal conduction, gas loss is largely driven by evaporation, a much faster process than ram pressure stripping. A bow shock and reverse shock propagate in these simulations, in the presence of conduction, although the bow shock is isothermal and not adiabatic, and the galaxy's core evaporates before the converged reverse shock propagates outward. 

For a more massive $2.8 \times 10^{12} \msun$ galaxy in an identical ICM wind, the process of evaporation and stripping is largely similar, although the evaporation time is $t = 550$ Myr. Qualitatively, the only significant difference in the evaporation process is that the galaxy's core does not evaporate before the expansion of the reverse shock. We also simulated the evaporation of a $2.8 \times 10^{11} \msun$ galaxy in a low density ICM of the same temperature ($\rho_{\rm ICM} = 2 \times 10^{-28}$ g cm$^{-3}$, $T_{\rm ICM} = 7.14 \times 10^7$ K). Due to a smaller heat flux, the evaporation time increases to $t \simeq 300$ Myr.

We therefore find that under realistic ICM and ISM conditions, isotropic thermal conduction efficiently removes the ISM at a significantly faster rate than ram pressure stripping. The removed ISM gas does not form tails, since it expands and evaporates reaching the temperature and density of the ICM before tails can form. In addition to the rapid removal of gas, thermal conduction suppresses the formation of KH instabilities. The existence of hot galactic coronae and stripped X-ray emitting tails in cluster galaxies therefore requires that thermal conduction is suppressed, likely due to ICM magnetic fields which in these conditions force thermal conduction to be effectively anisotropic, i.e., confine heat to flow along magnetic field lines only. 

\section*{Acknowledgments}
RV was supported by an NSF Astronomy and Astrophysics Postdoctoral Fellowship under award AST-1501374 and partially by the NASA Chandra theory award TM5-16008X. CLS was supported in part by NASA Chandra grants GO5-16131X and GO5-16146X and NASA XMM-Newton grants NNX15AG26G, NNX16AH23G, and NNX17AC69G. The simulations presented here were carried out on the Rivanna cluster at the University of Virginia. FLASH was developed largely by the DOE-supported ASC/Alliances Center for Astrophysical Thermonuclear Flashes at the University of Chicago. The figures in this paper were generated using the \texttt{yt} software package \citep{Turk11}. We thank the anonymous referee for useful comments that helped improve this paper.

\bibliography{ms}

\end{document}